\documentclass{article}
\usepackage{spconf,amsmath,graphicx}
\usepackage{color}
\usepackage{hyperref}
\usepackage{subcaption}
\usepackage{hyphenat}

\title{Parameterized Channel Normalization for Far-field Deep Speaker Verification}
\name{Xuechen Liu{$^1{}^,{}^2$}, Md Sahidullah{$^2$}, Tomi Kinnunen{$^1$}}
\address{
  {$^1$}School of Computing, University of Eastern Finland, Joensuu, Finland\\
  {$^2$}Universit\'{e} de Lorraine, CNRS, Inria, LORIA, F-54000, Nancy, France}

\begin{document}
%
\maketitle
\begin{abstract}
We address far-field speaker verification with deep neural network (DNN) based speaker embedding extractor, where  mismatch between enrollment and test data often comes from convolutive effects (e.g. room reverberation) and noise. To mitigate these effects, we focus on two parametric normalization methods: per-channel energy normalization (PCEN) and parameterized cepstral mean normalization (PCMN). Both methods contain differentiable parameters and thus can be conveniently integrated to, and jointly optimized with the DNN using automatic differentiation methods. We consider both fixed and trainable (data-driven) variants of each method. We evaluate the performance on Hi-MIA, a recent large-scale far-field speech corpus, with varied microphone and positional settings. Our methods outperform conventional mel filterbank features, with maximum of 33.5\% and 39.5\% relative improvement on equal error rate under matched microphone and mismatched microphone conditions, respectively. 
\end{abstract}
\begin{keywords}
acoustic feature extractor, channel normalization, spectrogram, far-field speaker verification.
\end{keywords}
\section{Introduction}
\label{sec:intro}

\emph{Automatic speaker verification} (ASV) \cite{asv2015} systems aim at verifying the speaker from input speech. ASV systems consist of three main components: acoustic feature extractor, speaker embedding extractor, and backend classifier. Thanks to advances in speaker embedding extraction based on deep neural networks (DNNs), ASV systems have improved substantially from conventional models such as i-vectors \cite{ivector}. This 
parallels advances 
in other speech tasks, such as automatic speech recognition (ASR) \cite{dnn_asr_2012} and keyword spotting (KWS) \cite{dnn_kws_2014}.

\begin{figure}[!t]
  \centering
  \centerline{\includegraphics[width=0.45\textwidth]{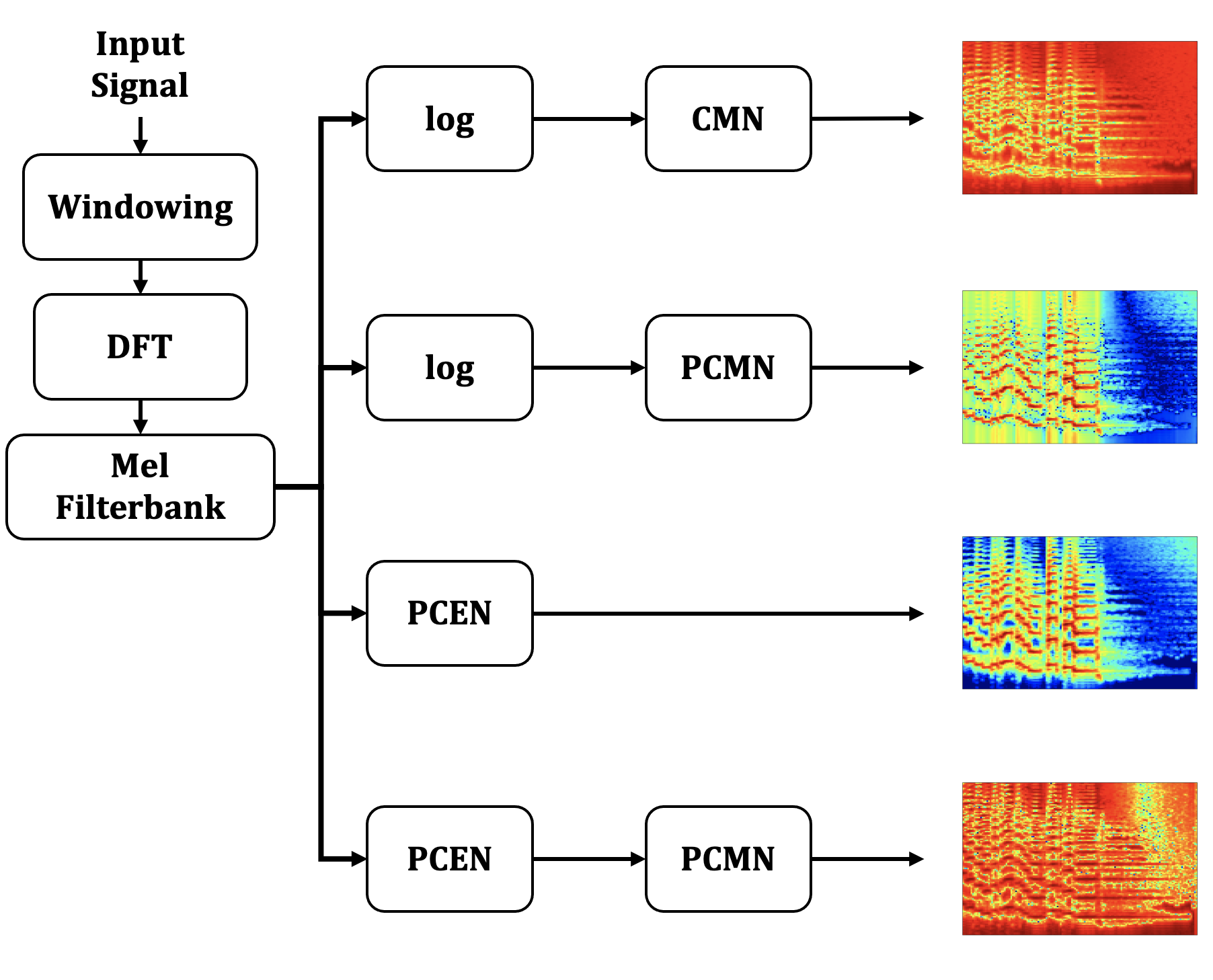}}
  \caption{Proposed feature extractor with channel normalization on mel filterbank, DFT: discrete Fourier transform, PCEN: per-channel energy normalization, CMN: cepstral mean normalization, PCMN: parametric cepstral mean normalization.}
  \label{fig:arch}
\end{figure}

While achieving promising performance under controlled conditions, ASV performance remains substantially low in far-field scenarios, where the user must be authenticated from a distance \cite{voices_corpus2018}. Far-field ASV is needed in multiple applications, from virtual assistants to teleconferencing. Given its evolving research value, dedicated datasets and benchmarks have been released. One example is \emph{Voices Obsecured in Complex Environmental Settings} (VOiCES) \cite{voices_corpus2018}, which formulates a multi-channel simulated distant ASV scenario. Another recent example is Hi-MIA \cite{himia}, a bilingual corpus (English and Mandarin) focused on smart home scenario with precise definition of microphone types and positions.

Far-field ASV is much more challenging compared to conventional ASV. The factors that impact recognition accuracy can be classified into two main categories: 1) \textbf{Environmental variations}, which includes natural room reverberation and additive noises. Common ways to tackle these include masking and de-reverberation techniques \cite{dereverb_asv_2019,far-field_asv_2019}, along with other speech enhancement techniques; 2) \textbf{Intrinsic variations} introduced by microphone array and speakers themselves. Earlier studies have addressed model-wise multi-channel training \cite{multichannel_ffasv_2019} and adding beamforming front-end \cite{voices_2020}.

In this work, we thus focus on improving acoustic feature extractor without increasing computational complexity substantially. Such efforts lead to decent progress in other speech processing tasks. One successful example is  \emph{power-normalized cepstral coefficients (PNCCs)} \cite{pncc}, whose efficacy has been demonstrated in speech recognition under various noisy conditions. Multi-taper MFCCs \cite{Tomi_multitaper2012} and linear predictive features target robust ASV, but the advantages demonstrated with conventional models \cite{lpcc_2014} do not necessarily generalize to DNN-based ASV \cite{Xuechen_feature2020}.

In this work, we extend upon earlier research by adopting two parametric methods that operate on time-frequency spectrogram to enhance feature robustness. The first method,  \emph{per-channel energy normalization} (PCEN) \cite{pcen_2017, pcen_2018}, replaces the commonly-used logarithmic compression. The second method, \emph{parameterized cepstral mean normalization} (PCMN) \cite{pcmn}, generalizes the widely-used parameter-free \emph{cepstral mean normalization} (CMN) by the inclusion of learnable normalization parameters. Even if PCEN and PCMN were addressed in other tasks, to the best of the authors' knowledge, they have not been extensively addressed in DNN-based ASV for far-field applications. Thus, the primary contribution of this work is integration (and joint training of) these compensation methods with modern DNN-based ASV systems. Finally, we also propose a novel channel normalization method that combines the two methods in cascaded fashion. Our experiments are conducted on the recent Hi-MIA corpus \cite{himia} with certain advantages elaborated below.


\begin{figure}[t]
  \centering
  \centerline{\includegraphics[width=0.4\textwidth]{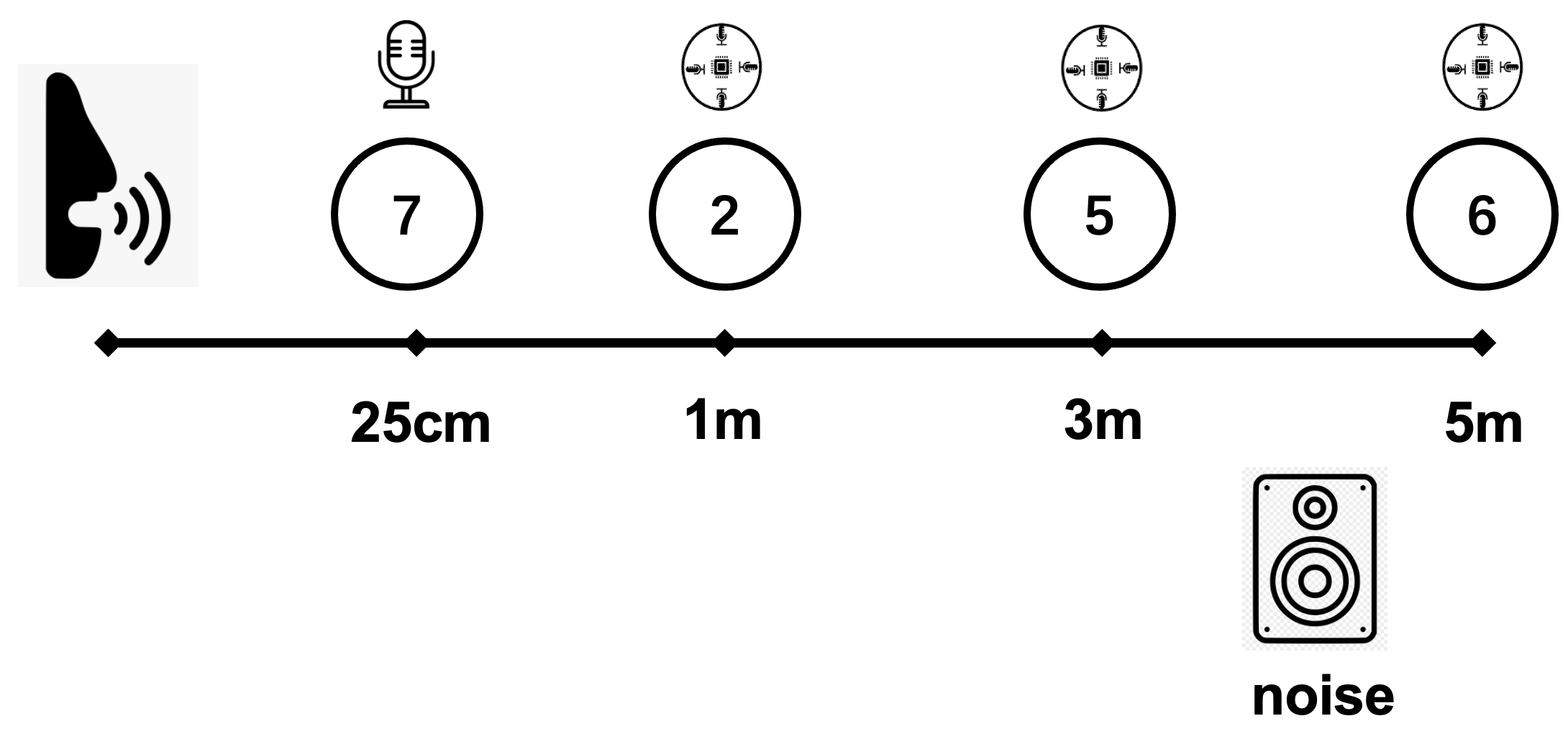}}
  \caption{Recording condition of Hi-MIA, adopted from \cite{himia}, excluding sources that are not included in our protocol. The distance between noise source and speaker is 4m.}
  \label{fig:himia}
\end{figure}

\section{Hi-MIA dataset for far-field speaker verification}
\label{sec:himia_data}

The work of this paper is conducted on Hi-MIA. It is a dataset which contains 1561 hours of audio from 340 speakers \cite{himia} recorded both under clean and smart home conditions in Mandarin and English. It consists of two subsets: \emph{AISHELL-wakeup} and \emph{AISHELL-2019B-eval}, which complement one another in terms of recording conditions. Speech was collected with both close-talking (clean) high-fidelity microphone and 16 types of distributed microphone arrays under real rooms, corresponding to smart home environments. This is one of the advantages of Hi-MIA compared to VOiCES \cite{voices_corpus2018}, where the source audios are replayed versions of pre-recorded audio. The positional information of different types of devices and noise resources is illustrated in Fig. \ref{fig:himia}. The publicly available part of the data contains recordings for the close-talking microphone (position 7) and the microphone arrays at positions 2, 5 and 6. For more details about gender and age distribution, refer to \cite{himia}. Hi-MIA shares similar recording conditions and contains overlapped speech with the evaluation data of the recent \emph{Far-Field Speaker Verification Challenge 2020} \cite{ffsvc2020}.

\begin{figure}[ht] 
  \begin{subfigure}[b]{0.5\linewidth}
    \centering
    \includegraphics[width=\linewidth]{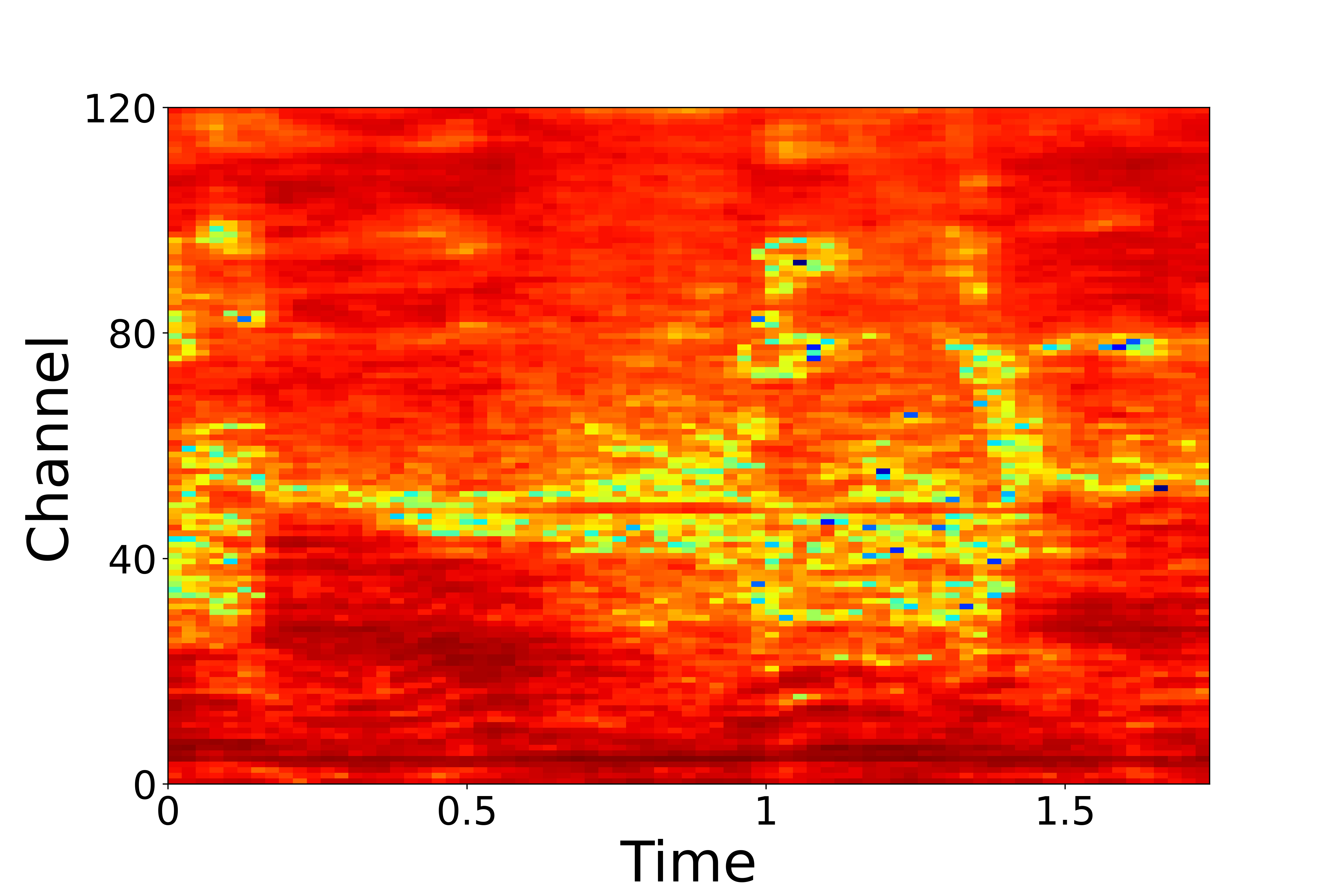} 
    \caption{log + CMN (baseline)} 
    \label{fig7:logcmn} 
  \end{subfigure}
  \begin{subfigure}[b]{0.5\linewidth}
    \centering
    \includegraphics[width=\linewidth]{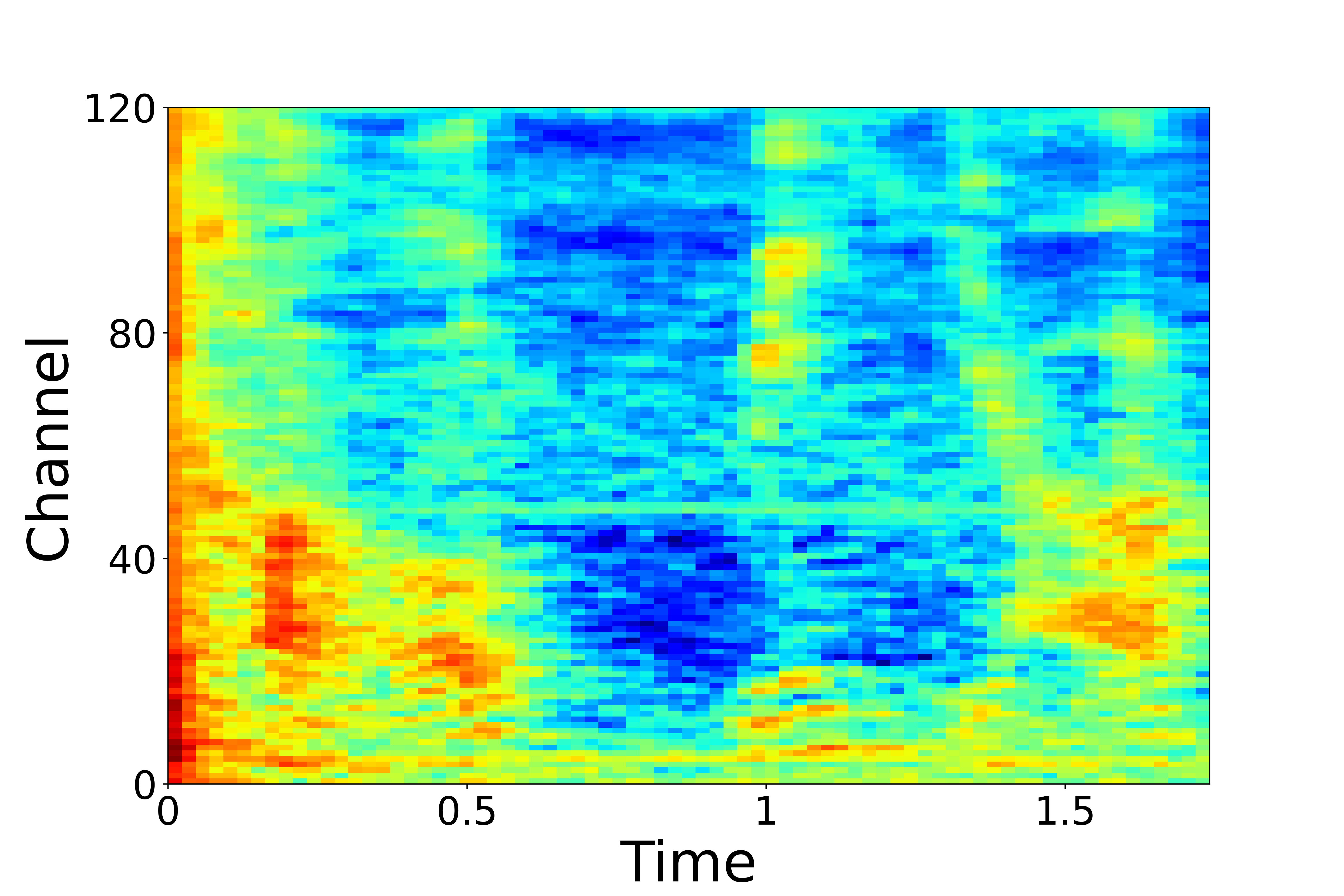} 
    \caption{PCEN} 
    \label{fig7:pcen} 
  \end{subfigure} 
  \begin{subfigure}[b]{0.5\linewidth}
    \centering
    \includegraphics[width=\linewidth]{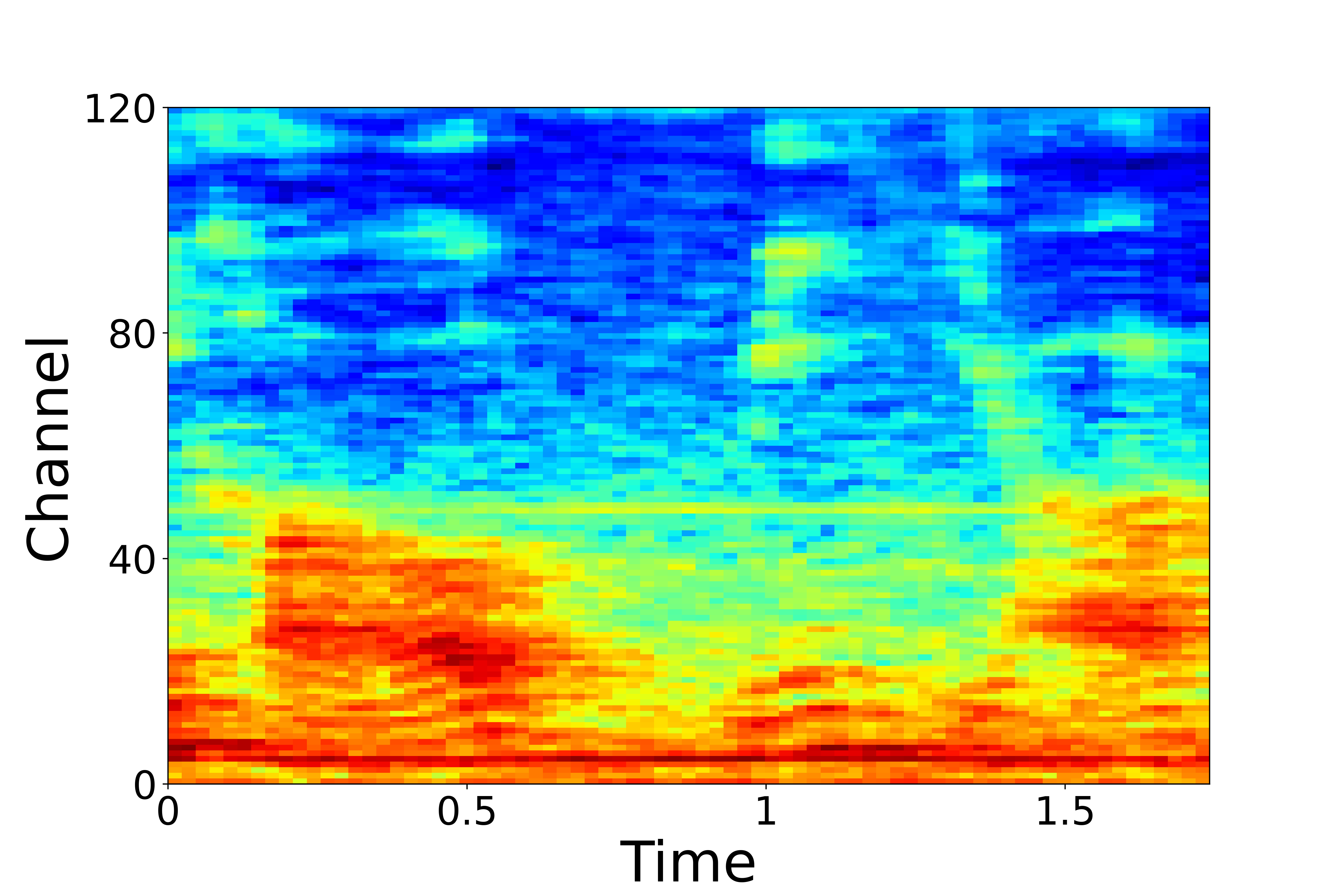} 
    \caption{PCMN} 
    \label{fig7:logpcmn}
  \end{subfigure}
  \begin{subfigure}[b]{0.5\linewidth}
    \centering
    \includegraphics[width=\linewidth]{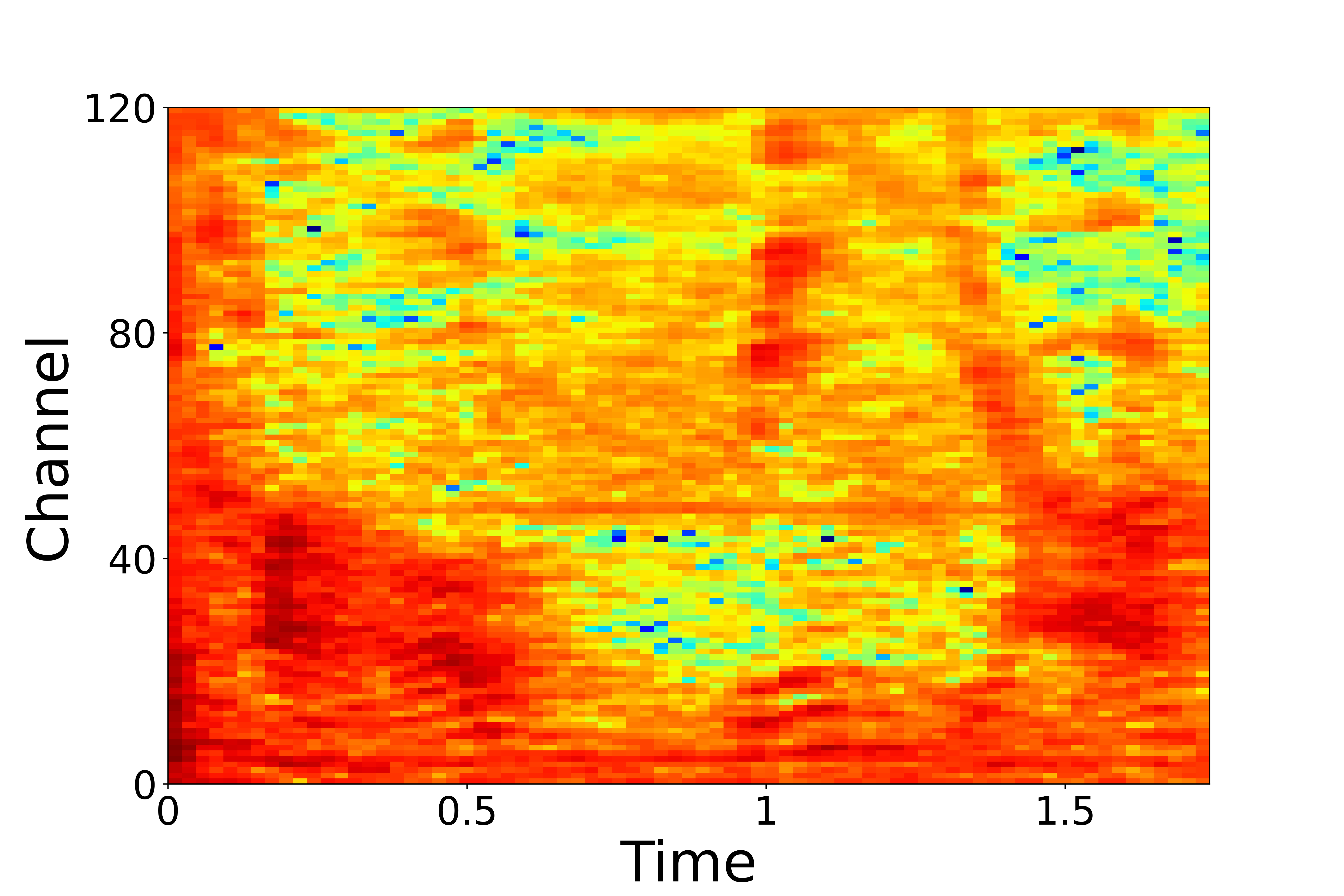} 
    \caption{PCEN + PCMN} 
    \label{fig7:pcenpcmn} 
  \end{subfigure}
  \caption{Different spectrogram representations on a speech utterance from Hi-MIA.}
  \label{fig:pcen_spec} 
\end{figure}

\section{Parameterized channel normalization}

\subsection{Per-channel Energy Normalization}
As a dynamic compression technique, PCEN \cite{pcen_2017} addresses the singularity problem of logarithmic compression at zero, which is non-trivial and not robust to environmental and loudness variations from speakers. With $t$ and $f$ being time and frequency indices, PCEN is formulated as:

\begin{align}
    \text{PCEN}[t,f] = \left(\frac{E[t,f]}{(M[t,f] + \epsilon)^{\alpha}} + \delta 
    \right)^{r} - \delta^{r}
\label{eq:pcen}
\end{align}

Here, $E[t,f]$ notes the input spectrogram energies. The PCEN operation in eq. \ref{eq:pcen} consists of two parts: \emph{automatic gain control} (AGC) and \emph{dynamic range compression} (DRC). AGC is represented by the term $G[t,f] = E[t,f]/(M[t,f] + \epsilon)^{\alpha}$, where $M[t,f] = (1-s)M[t-1,f] + sE[t,f]$ denotes temporally integrated energies, computed using a first-order infinite impulse response (IIR) filter with pre-set smoothing coefficient $s$. For all experiments in this paper, $s$ is the reciprocal to number of mel filters (40), following \cite{pcen_2017}. The main control parameter is $\alpha \in (0,1]$, which models the degree of compression. $\epsilon$ is a small number to avoid division by zero. Note that computation of $G[t,f]$ can be re-formulated to subtraction at logarithmic domain, followed by an exponential operation.

After obtaining the AGC-controlled energies, PCEN spectrogram is obtained by DRC, which is expressed as $(G[t,f] + \delta)^{r} - \delta^{r}$, where the positive bias term $\delta > 1$ and the exponent $r \in (0,1]$ are the main control parameters. They are designed to compress the loudness variations in the signal, reflecting earlier work on speech restoration \cite{speech_restoration}.

Fig. \ref{fig7:logcmn} and \ref{fig7:pcen} compare the baseline and PCEN normalized mel spectrograms on a speech utterance. PCEN performs extensive compression of the dynamic range in specific parts of the signal while keeping part of the pattern. On the other hand, it has enhancement effect on speech onsets, which may be helpful in improving robustness. More work on analyzing compression characteristics of PCEN can be found in \cite{pcen_2018}.

\subsection{Parametric Cepstral Mean Normalization}
Conventional CMN is a parameter-free and blind estimator which does not have interaction with other speech modules. It is expressed as below: 

\begin{align}
    \hat{X}_{t}[i] = X_{t}[i]-\mu_{t}[i]
\end{align}

where $t$ and $i$ are the time frame and feature indices, respectively. $\mu_{t}[i]=\sum_{m=t-N}^{t}X_{m}[i]/N$ stores cepstral mean values with a sliding window of length $N+1$. As a replacement, PCMN \cite{pcmn} is formulated as:

\begin{align}
    \hat{X}_{t}[i] = \beta[i]X_{t}[i]-(\alpha[i]\mu_{t}[i] + \mu_{0}[i])
\label{eq:pcmn}
\end{align}

where $\beta$, $\alpha$ and $\mu_{0}$ are the additional parameters. Compared to conventional CMN, this parameterized version allows the interaction between the normalizer and other learnable modules such as DNN speaker embedding extractor. It decides whether performing cepstral mean subtraction or not by varying the gain parameters $\alpha[i] \in [0,1]$.

Similar to PCEN, PCMN can also be effective on handling speech characteristics \cite{pcmn}. However, it may treat such information in a different way. Fig. \ref{fig7:logpcmn} shows the PCMN-processed mel spectrogram. Interestingly, the low-frequency speech components are generally preserved while high-frequency energies are 
subtracted to a lower level by PCMN. The compressed part becomes not as flat as observed from PCEN. This might be beneficial in preserving speech patterns as PCEN while creating more distinction and may also smear useful speech information. We take such a potential advantage and hypothesize that directly temporal modeling via PCMN can also lead to better ASV performance. 

\subsection{Trainable channel normalization}

While related parameters can be primarily set in hand-crafted fashion, both PCEN and PCMN have data-driven variants in their original works \cite{pcen_2017, pcmn}, making joint optimization via back propagation possible along with the DNN speaker embedding extractor.

For PCEN, as seen from eq. \ref{eq:pcen}, the parameters $\alpha, \delta, r$ are designed to be differentiable. Therefore, they can be generalized as frequency-dependent or even time-frequency dependent. Following \cite{pcen_2017}, we generalize them to be frequency-dependent and perform the joint optimization: $\boldsymbol{\alpha}=\alpha(f), \boldsymbol{\delta}=\delta(f),$ and $\boldsymbol{r}=r(f)$, where $f$ denotes frequency bin index.

For PCMN, from \cite{pcmn} the temporal dependency the method integrates can be leveraged by using a linear projection layer $\boldsymbol{W} \cdot \boldsymbol{Y}_{t} + \boldsymbol{b}$, where $\boldsymbol{Y}_{t} = [\boldsymbol{X}_{t-10}, ..., \boldsymbol{X}, ..., \boldsymbol{X}_{t+10}]$ are spliced cepstral input, $\boldsymbol{W} \sim (\boldsymbol{\alpha}, \boldsymbol{\beta})$ and $\boldsymbol{b}$ are corresponding weights and bias. The weights contain the frequency-dependent learnable values of $\alpha[i]$ and $\beta[i]$ and the bias can be regarded as a frequency-dependent variant of $\mu_{0}[i]$.

In the optimization of above trainable front-ends, we employ \emph{kernelized initialization}, where the parameters are not selected from a specific distribution such as normal one \cite{normal_distribution_2010}, but are migrated from common practical knowledge such as the workable hand-crafted counterparts. It has been applied and demonstrated to be effective for data-driven MFCCs \cite{learnable_mfcc2021}. Kernel initialization sets a starting point for further learning and adaptation via back-propagation. The exact values of related parameters are addressed in the next section.

\section{Experiments}

\subsection{Data}

We conducted all experiments on the Hi-MIA dataset. Training of the speaker embedding extractor was conducted using AISHELL-2 \cite{aishell2}, following the original Hi-MIA protocol \cite{himia}. The training data contains 1991 speakers recorded using an iOS device. The data was further augmented using room impulse response (RIR) \cite{rir} and noise sources from the MUSAN dataset \cite{musan}. 

Evaluation was conducted on the \emph{test} partition of Hi-MIA. Our protocol defines two trial sets and six trial lists in total, based on recording conditions illustrated in Fig. \ref{fig:himia}:
\begin{itemize}
    \item \emph{Matched microphone}, where the type of microphone between each enrollment and test utterances for each trial pair are the same. Enrollment data are recorded at position 2 while test utterances can originate from either 2, 5 or 6. This results in Ma-2, Ma-5, Ma-6 in Table \ref{tab:himia}.
    \item \emph{Mismatched microphone}, where the enrollment utterances are recorded using the close-talk microphone at position 7 while the test utterances are from the microphone arrays at position 2, 5 and 6. This results in Mis-2, Mis-5, Mis-6 in table \ref{tab:himia}.
\end{itemize}
The protocol of full trial lists are available as an open-sourced Kaldi\footnote{\href{https://github.com/kaldi-asr/kaldi/tree/master/egs/hi\_mia/v1}{https://github.com/kaldi-asr/kaldi/tree/master/egs/hi\_mia/v1}} recipe. 

\begin{table}[h]
    \centering
    \begin{tabular}{|c|c|}
    \hline
        Module & Values \\ \hline
         PCEN & $\alpha=0.98, \delta=2.0, r=0.5$ \cite{pcen_2017}  \\ \hline
         PCMN & $\beta=1.0, \alpha=0.5, \mu_{0}=0.0$ \\ \hline
    \end{tabular}
    \caption{Setting of parameter values for related modules for fixed front-ends. They are also used for kernel initialization if applicable.}
    \label{tab:parameters}
\end{table}

\begin{table}[h]
  \small
  \centering
  \begin{tabular}{|c|c|c|c|c|}
    \hline
    Trial ID & \#Enroll & \#Test & \#Target & \#Nontarget  \\ \hline
    Ma-2 & Arr, Pos. 2 & Arr, Pos. 2 & 35910 & 35250 \\ \hline
    Ma-5 & Arr, Pos. 2 & Arr, Pos. 5 & 34920 & 35175 \\ \hline
    Ma-6 & Arr, Pos. 2 & Arr, Pos. 6 & 34575 & 35175 \\ \hline
    Mis-2 & Mic, Pos. 7 & Arr, Pos. 2 & 34845 & 35010 \\ \hline
    Mis-5 & Mic, Pos. 7 & Arr, Pos. 5 & 35625 & 34920 \\ \hline
    Mis-6 & Mic, Pos. 7 & Arr, Pos. 6 & 34860 & 35655 \\ \hline
  \end{tabular}
\caption{Statistics for Hi-MIA sub-trials. Arr: Microphone array, Mic: close-talking microphone. Pos: position IDs where audios are recorded, referred from Fig. \ref{fig:himia}.}
\label{tab:himia}
\end{table}

\subsection{System Configuration}
\textbf{Front-ends}. For all ASV systems considered the number of mel filters was set to 40, which is same as the input feature dimension for speaker embedding network. We consider mel filterbank with logarithmic compression and CMN post processor as our baseline, as illustrated in Fig. \ref{fig:arch}. Additionally, we combine PCEN and PCMN, where PCEN replaces the logarithmic non-linearity. We refer to front-ends with fixed and trainable (adaptive) components, respectively, as \emph{fixed front-ends} and \emph{adaptive front-ends}. Details of exact parameter values for fixed front-ends and kernel initialization is shown in Table \ref{tab:parameters}. For kernel initialization, the values of vector-wise parameters are set as constant across all elements.

\textbf{Speaker embedding extractor}. Extended x-vector based on time-delayed neural network (E-TDNN) \cite{Snyder_etdnn_2019} is used as speaker embedding extractor. It is one of the descendants from \emph{x-vector} which have reached promising performance using mel cepstral features. We introduce two modifications from original configuration: 1) For pooling layer, we acquired attentive statistics pooling \cite{astats_pooling}; 2) For loss function we make use of additive margin softmax \cite{aam_softmax}. For inference, we extract 512-dimensional embedding vector for each utterance from the first fully-connected layer after the pooling layer. 

\textbf{Backend}. For all the experiments, we train corresponding probabilistic linear discriminant classifiers (PLDA) using the speaker embeddings from the \emph{train} partition provided by Hi-MIA open-sourced protocol. Embeddings are processed via mean subtraction, length normalization, and centering using a 200-dimensional linear discriminant analyzer (LDA) before being fed to the PLDA. 

\textbf{Evaluation}. Results are reported in terms of equal error rate (EER \%). We also provide detection error trade-off (DET) curves of fixed front-ends for analysis.

\begin{table*}[htbp]
  \centering
  \begin{tabular}{|c|c|ccc|ccc|}
    \hline
    & & \multicolumn{3}{|c|}{Hi-MIA matched mic} & \multicolumn{3}{|c|}{Hi-MIA mismatched mic} \\ \hline
    Non-linearity & Post norm. & Ma-2 & Ma-5 & Ma-6 & Mis-2 & Mis-5 & Mis-6 \\ \hline
    log & CMN & 3.65 & 7.43 & 8.51 & 8.16 & 11.01 & 12.84 \\ \hline
    PCEN & - & 4.05 & 6.93 & 6.87 & 6.45 & 10.7 & 12.34 \\ \hline
    log & PCMN & \textbf{3.27} & \textbf{5.56} & \textbf{5.66} & \textbf{4.35} & \textbf{6.66} & \textbf{9.23} \\ \hline
    PCEN & PCMN & 4.32 & 7.23 & 7.61 & 6.18 & 9.87 & 11.74 \\
    \hline
  \end{tabular}
\caption{EER (\%) results on Hi-MIA for fixed front-ends.}
\label{tab:nonadapt}
\end{table*}

\begin{table*}[htbp]
  \centering
  \begin{tabular}{|c|c|c|ccc|ccc|}
    \hline
    & & & \multicolumn{3}{|c|}{Hi-MIA matched mic} & \multicolumn{3}{|c|}{Hi-MIA mismatched mic} \\ \hline
    Non-linearity & Post norm. & Kernel init. & Ma-2 & Ma-5 & Ma-6 & Mis-2 & Mis-5 & Mis-6 \\ \hline
    aPCEN & - & no & 4.35 & 8.25 & 9.28 & 7.44 & 12.64 & 15.4 \\ \hline
    aPCEN & - & yes & \textbf{3.94} & 8.05 & 9.11 & \textbf{5.85} & 12.47 & 15.67  \\ \hline
    log & aPCMN & no & 4.45 & 8.85 & 9.14 & 8.94 & 13.24 & 14.58 \\ \hline
    log & aPCMN & yes & 4.31 & 7.43 & 8.9 & 6.17 & 11.59 & 16.09 \\ \hline
    aPCEN & aPCMN & no & 4.35 & 8.25 & 9.29 & 7.44 & 13.64 & 16.4 \\ \hline
    aPCEN & aPCMN & yes & 4.26 & 8.65 & 9.65 & 7.37 & 13.52 & 16.25 \\ \hline
    \hline
    aPCEN(no DRC) & - & yes & 4.17 & 8.55 & 8.97 & 7.07 & 13.02 & 16.08 \\ \hline
    aPCEN(no AGC) & - & yes & 4.60 & \textbf{6.66} & \textbf{7.48} & 6.34 & \textbf{9.90} & \textbf{10.95} \\ \hline
  \end{tabular}
\caption{EER (\%) results on Hi-MIA for adaptive front-ends, including ablation study for adaptive PCEN.}
\label{tab:adapt}
\end{table*}

\section{Results}

The results on fixed and adaptive front-ends are presented in Table \ref{tab:nonadapt} and \ref{tab:adapt}, respectively. Prefix 'a' added to the beginning of component indicates the adaptive variants.
Recall that the mismatched microphone scenario contains additional mismatch between high-fidelity close talking microphone and the microphone array. The enrollment utterances are always recorded with microphone distance of 25cm from the speaker. Therefore, it is expected that for test utterances at same position, corresponding EERs are generally higher. This is what we indeed observe in both scenarios. In the following, we discuss results for matched and mismatched scenarios one by one.

\subsection{Matched Microphone}

\textbf{Fixed front-ends}. All the proposed variants outperform the baseline when there is a large distance between the enrollment and test recordings (Ma-5, Ma-6). Lowest EERs come from log+PCMN, where the maximum relative improvement of 33.5\% over baseline comes from the furthest microphone position. Meanwhile, when the sources come from the same position, systems with PCEN do not outperform the baseline. 
Nonetheless, they do reach slightly lower EERs when the testing microphone moves from position 5 to 6. This indicates the potential of PCEN in normalizing room acoustic differences. Combining PCEN and PCMN gives slightly worse numbers than PCEN for all three trials. The comparative difference between PCEN and PCMN may be attributed to suboptimal parameter values: for computational reasons, we chose the values based on earlier recommendations from \cite{pcen_2018} and \cite{pcmn}. 

\textbf{Adaptive front-ends}. Somewhat disappointingly, results for the adaptive front-ends indicates generally worse performance compare to their fixed counterparts. The only system that returns satisfying performance for certain conditions is adaptive PCEN, which outperforms fixed PCEN by relatively 2.5\% in Ma-2. This indicates sensitivity of data-driven methods, especially there is mismatch between training and test data. There are meanwhile numbers of points can be improved related to engineering issues and domain adaptation on those parameter values, left as a future work. As a starting point, we see that kernelized initialization of the parameters gives slight relative improvement for most cases. Maximum improvement made by kernelized initialization is from aPCMN in Ma-5, by relatively 16\%. Such finding furthers the potential importance of hand-crafted knowledge to robust acoustic features.

\subsection{Mismatched Microphone}

\textbf{Fixed front-ends.} The improvement brought by fixed front-ends agree with ones in the matched microphone scenario. PCMN with logarithmic nonlinearity retains the lowest EERs overall, with a maximum relative improvement of 46.6\% in the Mis-2 condition. At the same time, for the two PCEN variants, better verification performance compared to the baseline on all cases is observed. Different from matched microphone condition, here the combination of PCEN and PCMN outperforms PCEN-only front-end, especially at the furthest microphone condition (Mis-6, relatively 4.8\%). 

\textbf{Adaptive front-ends.} For Mis-2 condition with 75cm distance between the close talking microphone and the arrays, most adaptive front-ends outperform the baseline. Lowest EER is obtained by aPCEN with kernel initialization in Mis-2, outperforming baseline by relatively 28.3\%. Nonetheless, for the other two cases, no adaptive front-end feature extractor gives lower EER than baseline and the performance gap between adaptive and fixed systems is noticeable. Also, while still giving relatively lower EER in Mis-5 by a maximum of 16.5\% (aPCMN), kernel initialization degrades the performance for all adaptive systems in Mis-6 condition.

\begin{figure}[ht]
  \centering
  \centerline{\includegraphics[width=0.5\textwidth]{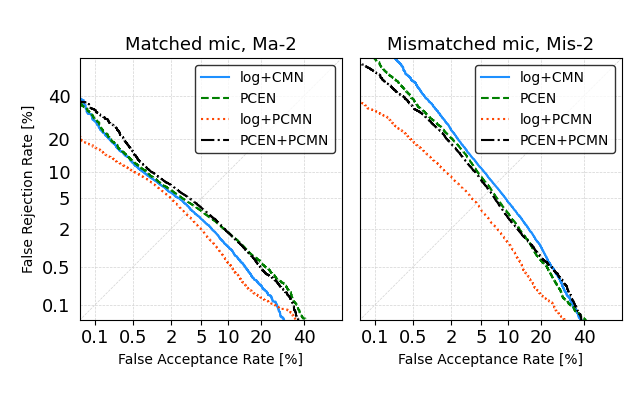}}
  \caption{DET curves for fixed front-ends.}
  \label{fig:det}
\end{figure}

\begin{figure}[ht]
  \centering
  \centerline{\includegraphics[width=0.55\textwidth]{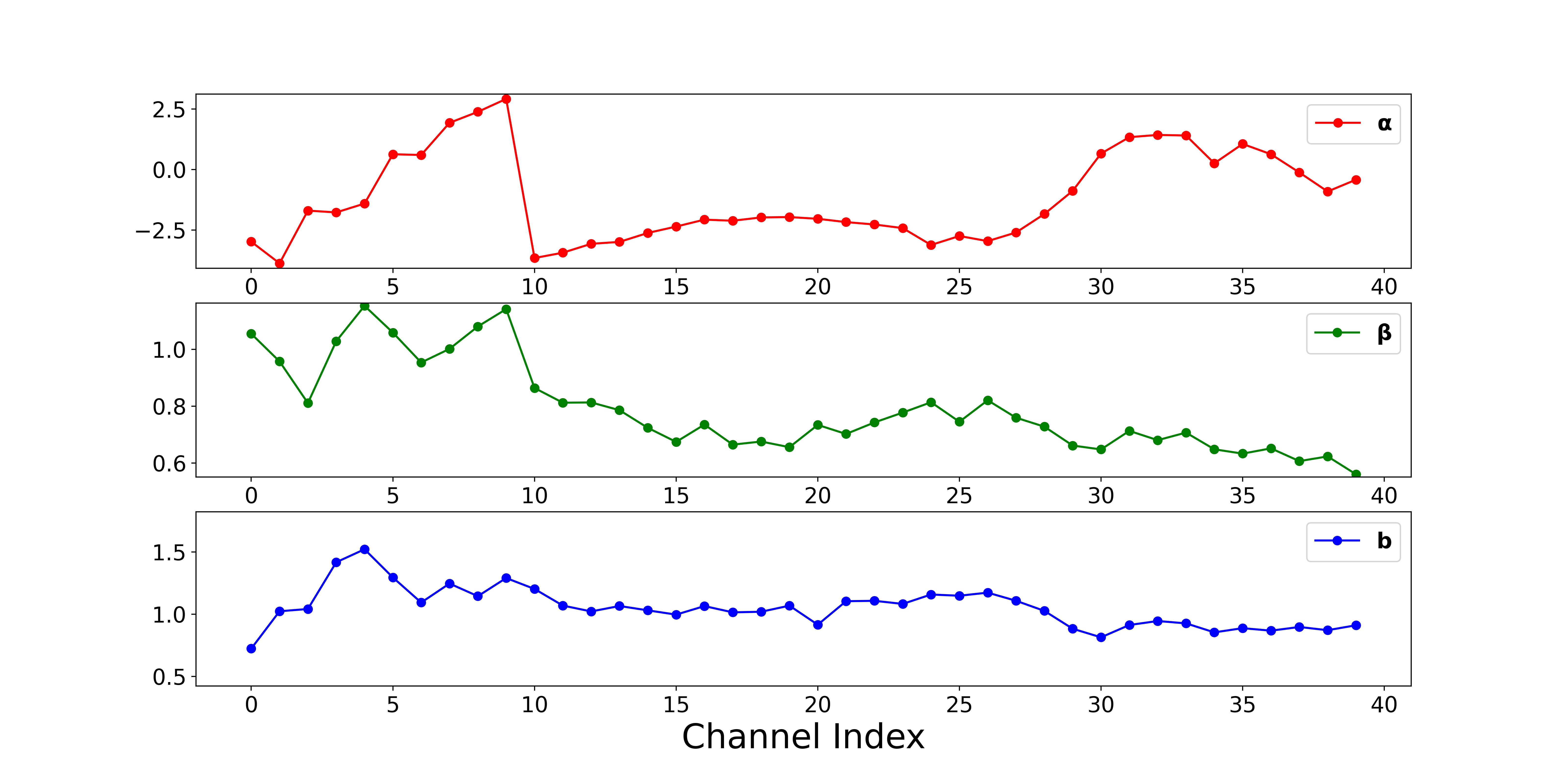}}
  \caption{Values of learnt weights $\boldsymbol{\alpha}$, $\boldsymbol{\beta}$ and bias $\boldsymbol{b}$ for system with log and aPCMN (with kernel initializaion). Number of channel indices is $i=40$ for all of them.}
  \label{fig:apcmn_params}
\end{figure}

\subsection{Analysis with DET plots}
DET curves of fixed front-ends for PCEN and PCMN on Ma-2 and Mis-2 are illustrated in Fig. \ref{fig:det}. The DET curves re-iterate what has been reflected in the EERs. Interestingly, however, when the false alarm rate gets high, performances of different front-ends approach to each other. Especially for the combination of logarithmic compression and PCMN: in fact, under Ma-2 it is outperformed by the baseline as can be noticed from the figure. Therefore, it may not be a good option for systems that are less strict on false alarms.

\subsection{Analysis with Adaptive Components}
\textbf{Ablation study on PCEN}. As noted earlier, PCEN consists of two main components: AGC and DRC. As our last experiment, we address their effect by removing one of them from the adaptive pipeline. The results are shown at the last two rows of of Table \ref{tab:adapt}. By removing DRC the filterbank energies are divided by its filtered variant without further compression. Compared to full adaptive PCEN, the results become worse. 
Meanwhile, by removing AGC, the energies are directly compressed without the division. In this case, performance of aPCEN is substantially improved and outperforms baseline in all conditions, except for Ma-2 with least mismatch between enrollment and test utterances. Results on Mis-6 indicates reversed gap between baseline and outperform it by relatively 14.7\% EER. Such observations may unveil the possible disadvantage cast by AGC and the potential benefits brought by DRC as a non-linearity.

\textbf{Learnt values of PCMN}. Finally, we show the weights and bias values of the best-performed system with adaptive PCMN involved in Fig. \ref{fig:apcmn_params}, where we combine logarithmic compression, adaptive PCMN and kernel initialization. As is evident, the learnt normalizers are different for each channel (unlike in the conventional non-parametric CMN). Further interpretation of the learnt normalization operations, and particularly their dependency on the training data, is deferred to a future work.



\section{Conclusion}
We addressed far-field speaker verification problem with mismatch conditions introduced by room reverberation and acoustic noise. We proposed new feature extractors to alleviate the negative impact due to these mismatches on verification performance by introducing two parameterized techniques on channel normalization: PCEN and PCMN. Our results on the recent Hi-MIA dataset confirm the efficacy of the introduced methods, especially for fixed PCMN. Our ablation study indicated the potential of DRC from PCEN.

In future research, we plan continue to explore both methods and their ingredients, especially compression parts such as DRC as a non-linearity itself and better integration on PCEN and PCMN with factorization, in order to compensate the high mismatch created by microphone and the factors.

\section{Acknowledgements}
This work was partially supported by Academy of Finland (project 309629) and Inria Nancy Grand Est.

\bibliographystyle{IEEEbib}
\bibliography{strings,refs}

\end{document}